\documentstyle[preprint,aps,eqsecnum]{revtex}

\newcommand\ignore[1]{}

\newcommand{\be}{\begin{equation}}
\newcommand{\ee}{\end{equation}}
\begin{document}

\draft
\title{Anisotropy and magnetism  of high temperature oxides superconductors }
\author{Jacques Friedel${}^1$ and
Mahito Kohmoto${}^2$}
\address{${}^1$Laboratoir\`{e} de Physique des Solides, Universit\'{e} Paris-Sud,
 Centre d'Orsay, 91405 Orsay Cedex, France, unit associated to the CNRS}
\address{${}^2$Institute for Solid State Physics, 
University of Tokyo, 7-22-1 Roppongi, Minato-ku, Tokyo, Japan}

\maketitle
\begin{abstract}
Phonon or electron mediated weak BCS attraction is enough to have high critical
temperature if a van Hove anomaly is at work. This could apply to electron doped
compounds and also to compounds with CuO$_2$ planes overdoped in holes, where $T_c$
decreases with increasing doping. If phonons dominate, it should lead to an
anisotropic but mainly $s$ superconductive gap, as observed recently in overdoped
LaSrCuO, and probably also in electron doped compounds. If electrons dominate, a
$d$ gap should develop as observed in a number of cases. In the underdoped range,
the observed decrease of $T_c$ with hole doping can be related in all cases to the
development of antiferromagnetic fluctuations which produces a magnetic pseudogap,
thus lowering the density of states at the Fermi level. The observed mainly $d$
superconductive gap then can be due to a prevalent superconductive
coupling through antiferromagnetic fluctuations; it could also possibly be
attributed to the same phonon coupling as in the overdoped range, now acting on
Bloch functions scattered in the magnetic pseudogap. More systematic studies of
superconductive gap anisotropy and of magnetic fluctuations would be in order.

\end{abstract}

\narrowtext

\section{Introduction}
Since the discovery of the perovskite copper oxides with superconductive critical
temperature $T_c$ between 30 and 40K\cite{BM},
the high $T_c$ copper oxides with parallel CuO$_2$ layers have been much
studied because of the hope, soon realised, of increasing $T_c$ still further and
also because such high value of $T_c$ did not fit a 'classical' BCS formula. Indeed
$T_c$ is then related to the Debye temperature $T_D$ by\cite{BCS}

\be
T_c\simeq T_Dexp(-1/\lambda) \label{bcstc},
\ee
where the exponential is very small for a weak relative coupling

\be
\lambda = Vn(E_F) \simeq V/E_F,
\ee
in which $V$ is the phonon mediated electron-electron coupling, $n(E_F )$ is the 
electron density of states, $E_F$ the Fermi energy. With $T_D \leq 300K$ 
and $\lambda \ll 1$, $T_c$ up to 25K should be very
exceptional. Classical corrections for larger phonon coupling $\lambda$ or for
effective Coulomb repulsion $\mu*$\cite{McMillan,ma,com} do not significantly alter
conclusions.

It is then natural to examine specific properties of HTSC:

--quasi two-dimensionality, with weakly coupled CuO$_2$ layers.

--antiferromagnetic(AF) fluctuations. The mother materials of HTSC are insulators
with AF order below $T_N$.

--anisotropic superconductive gaps. 

The phase diagram with doping and temperature  is indeed very rich, with the
antiferromagnetic ordered phase, the superconductive phases -- underdoped and
overdoped--, a pseudo AF gap region as well as a 'normal' metallic range as
shown schematically in Fig. 1 for the usual hole doped compounds. Electron doped
compounds show a similar succession of AF, superconductive and normal metallic
phases, with smaller critical temperatures.

Finally, in common with the classical BCS case, the superconductive state occurs
through condensation of cooper singlet pairing $({\bf k}\uparrow, -{\bf
k}\downarrow)$.

Our purpose is first to stress again that a usual BCS weak coupling scheme of
delocalised electrons is enough to explain the high value of  $T_c$ observed. The
observed anisotropy of the superconductive gap might tell us whether its coupling
is predominantly through electrons or through phonons. But the possible variation
of this anisotropy with doping must take into account the scattering of the Fermi
wave functions in the presence of AF fluctuations. It is then clear that the
nature of the superconductive coupling, whether by phonons or electrons, is more
directly related to the anisotropy of the gap in the  overdope range of  'normal'
metallic behavior.

\section{ CuO$_2$ layers }
We will use a  well-known
simplified picture restricted to positive holes of the $2p^6$ of the oxygen and of
the $3d^{10}$ of the copper, moving in a CuO$_2$ plane.
Only the orbitals O$_{2p}$ and Cu$_{3d}$ orbitals occupied by these holes are
considered.

\subsection{Electronic structure}
The three essential energies are 

$U$: the positive energy of  repulsion for two electrons on the same  site.

$\Delta$: the energy of promotion $3d \rightarrow 2p$ for a hole.

$t$: the absolute value of the transfer integral of $3d \leftrightarrow 2p$, 
which gives the frequency at which a positive hole  can transfer between a Cu and a
neighboring O. 

--In HTSC compounds, $X$ rays absorption data for the inner ${\em s}$ shells
of O show that the {\em 2p holes} exist in appreciable number for
 {\em undoped} as well as for doped samples\cite{friedel}. The same
conclusion can be drawn from the observed shift of the NMR line of Y in
YBaCuO\cite{alloul}. These fundamental observations are only compatible with
 
\be
 |\Delta|\leq  t.
\label{delta}
\ee
If we neglect for the moment the possible effect of $U$ in the AF compounds, the
positive holes are then in a band of {\em extended states} with a width $w$
of the order of $t$\cite{friedel}.

--The {\em 3d holes} are correlatively fewer than if all the holes were
concentrated on the Cu ions. They will be less than 1 on average on each copper in
the undoped samples, and their short range repulsion $U$ cannot lead to
any Verwey-Mott Coulomb localisation. In the absence of the magnetic effects
discussed later, the  holes should form, independently of doping, a fermi
liquid, with a Fermi surface and a band structure, responsible for metallic
conduction.

More precisely, if $U\gg w \simeq t$, the situation would recall  that of 
heavy fermion compounds, where $U$ acts mostly by increasing  the effective
mass in energy.

If $U \simeq w \simeq 2t$, the situation is analogous to that of
transition metals, where the essential  effects of correlation due to $U$
can be treated by perturbation\cite{fs}.

The position of the $X$ rays satellite emission line due to double excitation on Cu
in these compounds in fact compares with that observed in a metal like
Ni\cite{friedel}; this suggest that, although sizable, the effects of $U$ can be
usefully computed with a perturbing scheme, i.e. $U \simeq w$. The study of the 
limits of stability (in temperature and in doping) of the observed AF phase, as
discussed later, agree with this conclusion. This is what will be done in this
paper, although there is no essential difference in the metallic phase when $U \gg
w$.

\subsection{Approximate band structure [6]}
In the tight binding approximation used here, the Fermi level falls near the middle
of the antibonding CuO$_2$ band which couples the $3d_{x^2 - y^2}$ orbitals of Cu
and the
$2p_{\sigma}$ orbitals of O (Fig. 2). The one particle band structure should
depend on $t$, $t$' (transfer integrals in the $x$ and $y$ directions). Comparison
with more exact band structure calculations lead, for typical HTSC' like LaSrCuO
and YBaCuO to the following order of magnitude:

\begin{eqnarray}
\Delta &\simeq& 1eV, \nonumber \\
t &\simeq& 2eV, \\
|t-t'| &\simeq& 10^{-2}eV \; {\rm (in\;orthorhombic \; YBaCuO)} \nonumber ,
\end{eqnarray}
and thus the total band width of the  band  is

\be
w \simeq 5 eV .
\ee
This  leads to to an effective tight binding picture between
Cu atoms

\be
E_k \simeq -2t_{\parallel}\cos k_x a -2t'_{\parallel} \cos k_y b \, , \label{band}
\ee
with

\be
t_{\parallel} \simeq {t \over 2\sqrt{2}}\: ,
\ee
and
\be
w \simeq 8t_{\parallel}.
\ee

For undoped compounds, the Fermi level is then a square in the tetragonal
phase( $t'=t$) and very near to it in the orthorhombic one ($t' \neq t$). Doping
by electrons and by holes produce almost square Fermi surfaces with nearly
symmetric deviations from the undoped square surface (Fig. 3). This fundamental
symmetry between electron and hole dopings fits well the general symmetry of the
phase diagram observed for electron and for hole doped compounds.

The Fermi surface has the obvious property of exact or nearly exact nesting by a
translation ${\bf Q}$ equal to or nearly equal to the reciprocal lattice period
${\bf Q}_R$, at the origin of the AF instability discussed later. But the Fermi
level sits also near a strong peak in the density of states, which diverges
logarithmically at ($\pm \pi, 0$), ($0,\pm \pi$) for the undoped  compounds (Fig.
4). This  van Hove anomaly is characteristic of the (quasi) twodimensional
compounds, where it is much stronger than in more isotropic $3d$ compounds.

\subsection{High value of $T_c$ in the overdoped metal range}

We can reasonably neglect possible effects of AF thermal fluctuations in this
'normal' metallic range (Fig. 1). A high critical temperature $T_{\parallel}$ can
then be obtained in the mean field BCS approximation.

This can be seen assuming an isotropic coupling $V$ and an anisotropic $s$
gap\cite{hs,bl}. Then, if the Fermi level $E_F$ is near enough to a van Hove
anomaly of the density of states $n(E)$, the BCS equation

\be
{2\over V} = \int_{E_F-k_B T_D} ^{E_F+k_B T_D} \tanh({E-E_F \over 2k_B
T_{\parallel}})\, n(E) {dE\over {E-E_F}},  \label{gap}
\ee
leads to values of $T_{\parallel}$ much larger than (\ref{bcstc}), obtained with
$n(E)$ constant.

The maximum of $T_{\parallel}$ in this approximation occurs when the Fermi level
sits on the van Hove anomaly, thus for a half filled conduction band, at zero
doping. For large enough value of $T_D$, (\ref{gap}) then gives, from (\ref{band})

\be
k_B T_{\parallel} \simeq 2t_{\parallel}\exp\, (-\sqrt{ \pi t_{\parallel}/ V}) \,.
\label{2dtc}
\ee
This is {\em much} larger than (\ref{bcstc}), both because $\lambda$ is replaced
by $\sqrt{\lambda}$ in the exponential and because $k_B T_D$ is replaced by an
electron energy $t_{\parallel}$ in front of the exponential: the cutoff in
(\ref{gap}) is provided by the fast fall of the wings of the van Hove anomaly more
than by $T_D$.

With $t_{\parallel}\simeq 1 eV$, (\ref{2dtc}) then leads easily to large value of
of $T_{\parallel}$, larger than of those observed, even for small value of
$\lambda$. Indeed

\be
\lambda \simeq V/ 4t_{\parallel} \simeq 0.1 \; {\rm gives} \;
T_{\parallel}\simeq 1000K.
\ee

Two major corrections should appreciably reduce this upper limit\cite{friedel}:

--in the overdoped normal metal range discussed here ($c>c_M$, Fig. 1), the  Fermi
level is appreciably shifted from its optimum value at zero doping.

--because of the weak coupling between the electronic structures of neighboring
CuO$_2$ planes, superconductive fluctuations should reduce somewhat $T_c$ from
$T_{\parallel}$.

Very rough estimates of then two effects lead, in the optimum doping range $c
\simeq c_M$, to

\be
T_c \leq 100K.
\ee

Minor corrections affecting the core of the van Hove anomaly and discussed in
\cite{friedel} are not changing such estimate. The possibility of anisotropic gap,
discussed below, should increase $T_{\parallel}$, thus $T_c$ but little.

In conclusion {\em $T_c$ up to 100K are easily compatible, in the
overdoped metallic range, with a weak coupling schemes, as could be obtained by
standard phonon or electron couplings}. Even higher values of $T_c$ could be
obtained in similar compounds with no AF instability.

\section{antiferromagnetism and superconductivity}

\subsection{Covalent antiferromagnetism [6]}

In the model of delocalised electrons which follows
from (\ref{delta}), the only magnetism possible is a {\em  band magnetism}.
The  classical motor of such a  magnetism is, following Stoner and Lomer,  the
Coulomb interaction $U$ (except in the limit of infinite $U$'s, where one should
consider the exchange interaction
$J-tS$ ($J$ exchange, $S$ overlap, cf.\cite{zr}) and/or direct transfers between O
$2p$ orbitals (Barisic, priv. comm.).

In fact, if  $U$ is treated in perturbation, one  expects  for
the {\em undoped } compounds, an antiferromagnetism with a  wave
vector 
${\bf Q}_R$ which is the side of the square, Fig. 3; and the value obtained for
$T_N$, including corrections for thermal fluctuations, is compatible with $U
\leq w$. Because only a fraction of holes is found on the Cu, of the order of
$1/2$ if $|\Delta| \leq w$, we can interpret in this way the observed magnetic
moments; quantum fluctuations of these moments are expected to be weak because, in
this description, each atomic moment on a copper ion interacts in perturbation
with many others, through the long range spin oscillation it produces.

The AF order leads to the opening of  an antiferromagnetic gap around the Fermi
level (Fig. 5), leading to a {\em band insulator} in the AF state. For $T$
slightly larger than $T_N$, we expect strong AF fluctuations, inducing an effect
of  {\em antiferromagnetic 'pseudogap'}, as described initially by Mott in
another context: without vanishing completely, the density of states should
decrease around the Fermi level $E_F$, between rounded bonding B (i.e.
antiferromagnetic) and antibonding AB (i.e. ferromagnetic) peaks, as pictured in
Fig. 5. For strong enough fluctuations, i.e. near enough to $T_N$, one expects a
localisation of  all the states in the pseudogap, thus an {\em insulator
by disorder}, {\`a} la Anderson.

For {\em weakly doped} compounds, the observed AF order preserves a
periodicity coherent with that of the crystal: this can be expected if there is
interference of spin density waves on the CuO$_2$, an ordinary phenomenon, to be
expected here especially by the fact that the faces of the square, Fig. 3, nest
exactly by the translation ${\bf Q}_R$  (cf. Appendix A). From the fact that 
${\bf Q}_R$ does not superpose the opposite faces of the Fermi surface in weakly
doped samples, a rapid decrease of $T_N$ is expected with increasing doping. This
is just what is observed (Fig. 1), and the value of the maximum doping of the AF
phase corresponds again to a modest value of $U \simeq w$.

Finally in the {\em more strongly doped} compounds,  the intensity of
the AF fluctuations decrease with increasing   temperature or doping. The limit of
notable fluctuations  is marked by $T_f$ in Fig. 1: they disappear in fact beyond
the peak of superconductivity $T_c (c_M)$, by smoothly joining $T_c$ around $c_t$
slightly larger than than $c_M$. Depending on the compound considered, the
average wave vector of these fluctuations remains at ${\bf Q}_R$ (YBaCuO) or
varies with doping like the nesting vector ${\bf Q}$ of the Fermi surface (Fig. 3,
LaSrCuO): such variation, in agreement with the Lomer criterion, confirms the role
of the Fermi surface in a weakly perturbed band structure. One expects also, below
$T_f$, the appearance of an AF pseudogap (Fig. 5), with a density of states at the
Fermi level decreasing with temperature and with doping. this phenomenon this
phenomenon first observed by NMR\cite{alloul,yasuoka,takigawa} in underdoped
YBaCuO, i.e. for $c < c_t$, Fig. 1 (by study of both the Knight shift and the
relaxation rate of Y, little sensitive to magnetic effects). Note that, in this
region,  the magnetic fluctuations seem to be too small to  localise strongly by
disorder the Fermi electrons.

\subsection{Superconductivity and density of states at the Fermi level [6]} 
In reducing the density of
states at the Fermi level, the AF pseudogap lowers  the critical superconductive
temperature $T_c$. This effect is the more important the lower the doping; it is
prolonged continuously in the region of AF gap, with vanishing density of states
and no $T_c$ at all. This explains naturally the rapid decrease of $T_c$ with
doping in the underdoped range ($c< c_M$, Fig.1); this effect is independent of the
nature of the coupling, if it is of the BCS type\cite{friedel}; it has recently
been taken
 into account in models of superconductive coupling through AF
waves\cite{pines2}.

It must be stressed that many authors have wanted to see in  the pseudogap 
appearing at $T_f$   an effect of superconductive fluctuations. It seems  a
priori not very  reasonable to try to explain by the same physical phenomenon  
 two curves, $T_f(c)$ and $T_c(c)$ so different in behavior in 
the phase diagram. One can  also remark  that  $T_f (c)$ varies as expected
 for the AF fluctuations, as observed e.g. by neutron scattering.
Finally the superconductive gap as measured by the Andreev reflection varies
quasiparabolically with $c$, as $T_c(c)$; but the gap observed by normal  tunnel
effect  or  optically  varies like $T_f(c)$, as  expected for a gap due to 
AF fluctuations\cite{deutscher}.

If this interpretation is correct, the AF fluctuations are much more developed
than the superconductive ones in the underdoped range, at least in the compounds
discussed here such as YBaCuO or LaSrCuO. This might at least partly be due to a
weakness of the magnetic interplane couplings compared with the superconductive
interplane couplings. Other HTSC compounds might deviate from the phase diagram
sketched Fig. 1.

\subsection{Nature of superconductive couplings }
In the  BCS type of approach used here, two types of couplings can
  be dominating: attractive phonons,  or repulsive electrons.
As in other compounds, the observation of a possible anisotropy of the
superconductive gap and a study of its form $\Delta_{\bf k}$ can give some
indication on the nature of the couplings $V_{{\bf k},{\bf k'}}$, as indicated by
the BCS gap equation

\be
\Delta_{\bf k} = -\sum_{\bf k'} \frac{V_{{\bf k}, {\bf k'}} \Delta_{\bf k'}}
{2\sqrt{\Delta_{\bf k'}^2 + \varepsilon_{\bf k'}^2}} \tanh{\frac{\sqrt{\Delta_{\bf
k'}^2 +\varepsilon_{\bf k'}^2}}{{ 2k_B T}}} .
\label{gapeq}
\ee
where $\varepsilon_{\bf k'}$ is the one-particle energy measured from the Fermi
level.

Following general arguments recalled in Appendix B, it is usually agreed that an
$s$ gap (Fig. 6a) or an anisotropic $s$ one (e.g. $s+g$, Fig. 6b) are the signature
of phonon attractive couplings, while a $d$ gap (Fig. 6c) or a $d+s$ one (Fig. 6d)
are due to predominantly electron-electron coupling, as due to AF waves for
instance.

Many observations have in fact indicated the presence of $d$ gaps in many HTSC
compounds, thus favoring the idea of coupling through AF
waves\cite{moriya,pines}. This would not be in disagreement with the general
conclusion of this paper that a simple BCS weak coupling scheme could apply to
HTSC compounds. In such a scheme, the exact role
of the van Hove anomaly would have to be studied, although it was implicitly taken
into account into computations\cite{pines}.


Conclusions from experiments have however to be analysed carefully. Thus

--electron doped compounds show few signs of $d$ gap symmetry, even if they have
admittedly lower values of $T_c$ than the corresponding hole doped ones. (However, 
there is a report on the existence of $d$ gap\cite{maeda}.)

--recent measurements by Deutscher {\it et al.}\cite{deutscher2} using Andreev
reflection techniques, show for LaSrCuO a $d$ gap for underdoped samples ($c<c_M$,
Fig.1), but an anisotropic $s$ gap for optimum ($c \simeq c_M$) and for overdoped
sample ($c > c_M$). These measurements for anisotropic $s$ gaps can indeed be
fitted with computations by Bouvier and Bok\cite{bok} who, following
Abrikosov\cite{abrikosov}, use a coupling by phonons screened in two dimensions.
Deutscher's results agree with Tsuei {\it et al.}'s\cite{tsuei} on the
quantification of vortices in grain boundaries giving a $d$ gap in YBaCuO ($c \leq
c_M$), although a $d+s$ gap is more likely near $c_M$\cite{beal}. Results by Tsuei
and others on a number of HTSC compounds seem generally to obtain a $d$ gap, but
probably mostly on samples where an overdoping is not sure, either in the grains
or in the grain boundaries.

It seems therefore that sizable $T_c$'s are compatible with observed (anisotropic)
$s$ gaps and phonon coupling.. And Eq. (\ref {gapeq}) then leads to such values of
$T_c$, due to the fact that the van Hove anomalies are near the maximum of the gap
in
${\bf k}$ space. This is a priori what is to be expected in the overdoped normal
metal range.

For the underdoped range ($c < c_M$), the situation might be more complex. It
might well be that a strong AF instability leads to a $d$ gap. In compounds such
as LaSrCuO with $s$ gaps for $c \geq c_M$, one must then go, on reducing doping,
through intermediary situations such as Abrikosov's (\cite{abrikosov}, Fig. 6e)
and B\'{e}al and Maki's (\cite{beal}, Fig. 6d) where either phonon or AF waves are
respectively predominant, but both types of couplings are sizable.

The existence of
an AF pseudogap for underdoped samples suggests however another interpretation of
$d$ gaps. In the case where the AF pseudogap is larger than the superconductive
gap, i.e. for large enough underdopings, electronic states coupled by BCS all
belong to the AF pseudogap, between the bonding B and the antibonding AB peaks,
Fig.6. Each electronic state is then close to a linear combination of the type

\be
|{\bf K} > = \alpha_{\bf k}|{\bf k}> + \beta_{\bf k} |{\bf k}-{\bf Q}>,
\ee
where $|\alpha_{\bf k}|\simeq |\beta _{\bf k}|$. The $d$ anisotropy of the  gap,
due to that of $V_{{\bf k}, {\bf k'}}$ can then come from that of the
potential of interaction $V_{{\bf k}, {\bf k'}}$.
More precisely a coupling $V_{{\bf k}, {\bf k'}}$ as described in
\cite{abrikosov,bok} (with $V_{{\bf k}, {\bf k'}} < 0$ and the maximum of
$|V_{{\bf k}, {\bf k'}}|$ for  ${\bf k}\simeq {\bf k'}$) leads to a $d$ gap if,
near the Fermi level, $\alpha_{\bf k}\beta _{\bf k} < 0$ on the average. This is
expected in hole doped YBaCuO, where ${\bf Q} = {\bf Q}_R > 2k_F$: the Fermi level
is then in the lower range of energy in the pseudogap, nearer to B (where
$\beta_{\bf k} \simeq -\alpha _{\bf k}$) than to AB (where $\beta_{\bf k} \simeq
\alpha _{\bf k}$). In a compound such as LaSrCuO, where ${\bf Q}$ is less than
${\bf Q}_R$, Appendix A shows that the same situation of Fermi level near the
lower edge of the pseudogap might apply, with the same result of $d$ gaps from
phonon coupling.

\section{conclusions}

Experimental evidence shows that, at least in compounds such as LaSrCuO and
YBaCuO, a delocalised electrons scheme with weak correlations applies to undoped
as well as to doped HTSC compounds of the cuprate family; and it would be of
interest to extend these $X$ rays and NMR studies, when possible, to other
compounds of the same family. It is also clear that, within such scheme, a weak
BCS coupling is enough to explain high value of $T_c$ if one takes into account
the presence of (quasi) 2d van Hove anomalies. Exact estimates are however made
difficult by the problems of interplane couplings and 2d superconductive
fluctuations.

The high anisotropy of the superconductive gap $\Delta$ explains the large
deviations of the observed ratios of $\Delta /T_c$ from what is expected from
the classical isotropic BCS equations. The anisotropic $s$ gaps observed in
electron doped compounds and, more recently, in optimally and overdoped
LaSrCuO ($c \simeq c_M$) are compatible with a classical phonon coupling. The $d$
gaps observed in other overdoped compounds can be due to electron coupling; they
would then be a signature of larger electron correlations.

The $d$ gap observed in underdoped LaSrCuO, YBaCuO and a number of the
other compounds might be due also to electron couplings which could take the form
of coupling by AF waves, especially present in underdoped samples. The AF wave
pseudogap, observed in the underdoped samples, decreases the density of states at
Fermi level and thus
$T_c$ when it develops, that is for decreasing dopings. This last effect is
independent of the nature of superconductive coupling; and indeed the presence of
AF pseudogap might possibly explain a phonon coupling leading to $d$ gap, owing to
the special symmetry of the Fermi electrons in the AF pseudogap: this might
explain the change of symmetry from $d$ to $s$ with increasing dopings in a
compound such as LaSrCuO. Finally this change of gap symmetry with doping, which
occurs near maximum of $T_c$, must proceed continuously, in a way that has not
been studied completely.

More experiments would be very valuable in this content, especially in three
domains:

--anisotropy of the superconductive gap versus doping for hole doped samples in the
overdoped and optimum ranges. And also for electron doped sample such as
\mbox{Nd$_{2-x}$Ce$_2$ CuO$_4$}, the doping dependence of the gap anisotropy might
shed a light on the essence of HTSC.

--more systematic relation between AF fluctuations and transport properties in
undoped as well as underdoped compounds, in electron doped as well as hole doped
compounds.

-- more precise properties of phonons in these materials have to be considered. If
only intraplane pairings are envisaged, there are six relevant modes of lattice
vibrations (there are three atoms CuO$_2$ in a unit cell). Two of them are
acoustic and the rest are optical. The considerable complexity of
these phonons\cite{egami}, especially optical ones, should be taken into account.
Certainly the simple models considered in \cite{abrikosov,bok} have to be
reexamined.\\

\begin {center}
{\large{\bf Acknowledgments}} 
\end {center}

It is a pleasure to thank B. Barisic, J. Bok, G. Deutscher, T. Egami, A. Maeda, K.
Maki, Y. Matsuda, Y. Petrov, D. Pines, J. Shiraishi, M. Takigawa, and K. Ueda for
useful discussions.\\

\appendix
\section{Stability of AF phases}

The band contribution to the  stability of an AF gap or pseudogap, as pictured
Fig. 5, is

\be
\delta E =  \int ^{E_F} n(E)E dE - \int ^{E_F^0} n_0 EdE,
\ee
if $E_F$ and $E_F^0$ are the Fermi levels in presence and in absence of the gap
respectively. The Coulomb correction in $U$ to this Hartree approximation is known
not to alter qualitatively the conclusions of this typed of analysis.

The average number of electrons per CuO$_2$ in a plane is 

\be
z = 1-c = \int ^{E_F} n(E) dE - \int ^{E_F^0} n_0 dE.
\ee
This gives

\begin{eqnarray}
\frac{d\delta E}{dz} &=& E_F - E_F^0,\\
\frac{d^2\delta E}{dz^2} &=& \frac{1 }{n(E_F )} - \frac{1}{n_0}.
\end{eqnarray}

It is then easy to check that, for a gap due to long range AF order at ${\bf Q} =
{\bf Q}_R$, $\delta E({\bf Q}_R, c)$ has a minimum at zero doping (Fig. A), with
a break in the slope $d\delta E /dc$ related to the gap $E_{AB} - E_B$ (Fig.
A), thus to $U$. The symmetry of $\delta E({\bf Q}_R,c)$ is related to the
symmetry of $n(E)$, itself due to the quasi one dimension of $E_{\bf k}$ near zero
doping (Fig. 3). For an AF order at ${\bf Q} < {\bf Q}_R $, the gap of $n(E)$ is
expected to be smaller, thus also the stability of the AF phase; the increasing
twodimensional nature of $E_{\bf k}$ produces, for a given ${\bf Q}$, a stability
$\delta E({\bf Q}, c)$ less marked and more asymmetrical, which can be reasonably
expected to be above $\delta E({\bf Q}_R, c)$ in most of the region of stability
of the AF order at ${\bf Q}_R$. This situation, pictured Fig A, could explain that
the AF phases with ${\bf Q} < {\bf Q}_R$ could only appear, at 0K, outside the
region of stability of the AF order at ${\bf Q}_R$.

In the fluctuating regimes at finite $T$'s, the pseudogap which replaces the gap
(Fig. 5) has the effect of diminishing the stability $\delta E ({\bf Q}, c)$, and
of rounding off the angle at the minimum of $\delta E$.

It seems that, for YBaCuO, the commensurate AF phase at ${\bf Q}_R$ is always the
most stable (case  ${\bf Q}$, Fig. A); in LaSrCuO, incommensurate AF phases
appear above a finite hole doping (case ${\bf Q}'$, Fig. A); because their
stability decreases fast with decreasing ${\bf Q}'$, they should appear
systematically with a hole doping $c$' slightly larger than that for the minimum of
$\delta E({\bf Q}', c)$: Thus, in Fig. A, the phase ${\bf Q}'(c)$ shroud  only
appear for hole concentrations larger than $c'$, where a common tangent touches
${\bf Q}_R (c)$ and {\bf Q}'(c): to AB, Fig. 5.  

\section{Symmetry of gap}
The interactions $V_{{\bf k}, {\bf k'}}$ obviously have the space group symmetry of
the square (dihedral symmetry). The group elements $T$ are unity; rotations by
$\pi/2$, $\pi$, and
$3\pi/2$; and reflections about $k_x$ and $k_y$ axes and the lines $k_y=
k_x$, and $k_y = -k_x$. Then one has

\be
V_{T{\bf k}, T{\bf k}'} = V_{{\bf k}, {\bf k'}}.
\ee
The gap $\Delta({\bf k})$ created by the interaction $V_{{\bf k}, {\bf k'}}$  does
not have to have the same symmetry, but it is significantly influenced.

All the reflections and a rotation by $\pi$ have the property $T^2 =$ unity.
Therefore even if the gap is not invariant with respect to these transformations
there is a possibility that $\Delta(T{\bf k})=-\Delta({\bf k})$. We call this
'odd parity' and for $\Delta(T{\bf k}) = \Delta({\bf k})$ 'even parity'.

In order to consider the symmetry  of $\Delta({\bf k})$, let us divide
the first Brillouin zone into eight octants ( Fig. B). For the singlet pairing
one has $\Delta(-{\bf k}) = \Delta({\bf k})$ (the symmetry of  rotation by $\pi$
for $V_{{\bf k}, {\bf k'}}$). Thus it is enough to consider the gaps in the four
regions in 
$k_y > 0$, $\Delta_1$, $\Delta_2$, $\Delta_3$, and $\Delta_4$ (Fig. B). The
four directions $ \pm k_x$ and $\pm k_y$ have the van Hove anomalies. The gap will
be large in these direction due to the enhanced density of states. Therefore we
have even parity for the reflections with respect to $k_x$ and $k_y$ axes. (Note
that if one has odd parity, $\Delta = 0$ for the  directions of the van Hove
anomaly.) 

Finally the parity with respect to the lines $k_y = \pm k_x$ will determine the
full symmetry of the gap.

i) even parity : isotropic $s$ (Fig. 6(a)),  $s+g(\ell = 4)$  (Fig. 6(b)) or
something else.

ii) odd parity: $d$ symmetry (Fig. 6(c)) 

For attractive interactions $V_{{\bf k}, {\bf k'}}$ Eq. (\ref{gapeq}) shows that
the optimum value of  $\Delta_{\bf k}$  has necessary a constant sign, and the
even parity follows. For phonon couplings, where $|V_{{\bf k}, {\bf k'}}|$ is
maximum for ${\bf k} = {\bf k'}$\cite{abrikosov,bok}, the van Hove anomalies lead
to a very anisotropic $s$ gap, Fig.6b.

For repulsive interactions, Eq.(\ref{gapeq}) requires the odd parity. For
couplings by AF fluctuations, $V_{{\bf k}, {\bf k'}}$ has a large peak at ${\bf
k}'\simeq {\bf k}-{\bf Q}_R $ (Fig. B); this leads to odd parity about the line
$k_y = k_x$ and $k_y =- k_x$, thus to a $d$ gap, Fig. 6c.

\begin{figure}

Fig. 1 
Schematic phase diagram of hole doped oxides. AF: antiferromagnetically
ordered phase; $T_c$: superconductive transition temperature; $T_f$: below this
temperature the AF fluctuations are notable;  $S$:
superconductive phase;
$c_M$: optimal doping; $c_t$: doping for disappearance of AF pseudogap.

Fig. 2
Cu $3d$ and O $2p$ orbitals and corresponding transfer integrals $t$, $t$' in a 
CuO$_2$ plane.

Fig. 3
Fermi surface.

Fig. 4
Density of states $n(E)$ and van Hove anomaly.

Fig. 5
Symmetry of the anisotropic gap: a. $s$ symmetry; b. 'anisotropic $s$'
symmetry; c. $d$ symmetry; d. $d$ + $s$ symmetry; e. very anisotropic $s$ gap.

Fig. 6
Gap ( $T < T_N$) and antiferromagnetic pseudogap( $T > T_N$).

Fig. A
Stability $\delta E$ of commensurate(${\bf Q}_R $) and incommensurate (${\bf Q}$)
AF phases at $OK$ versus hole doping $c$.

Fig. B
$\Delta$'s near half-filling.

\end{figure}

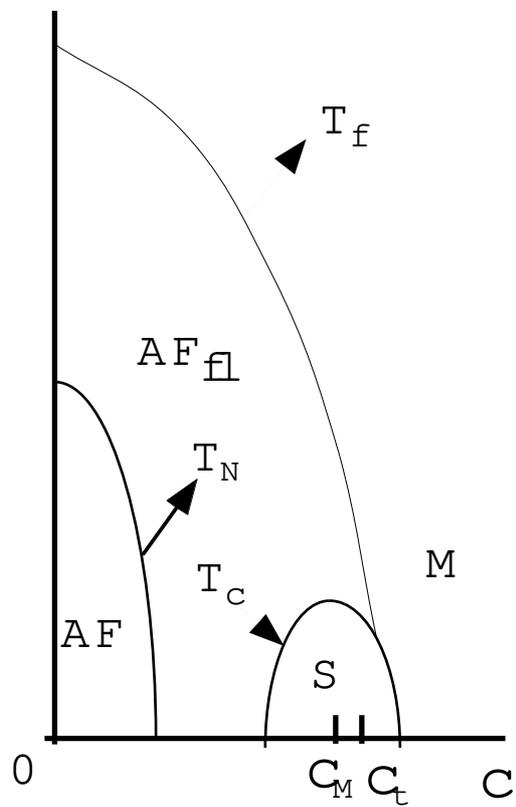
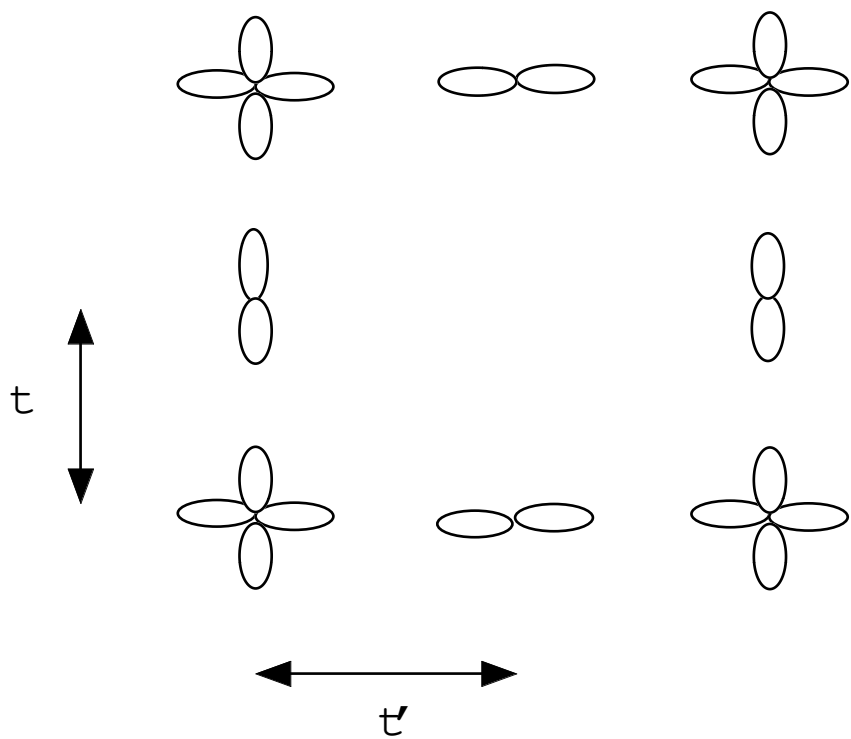
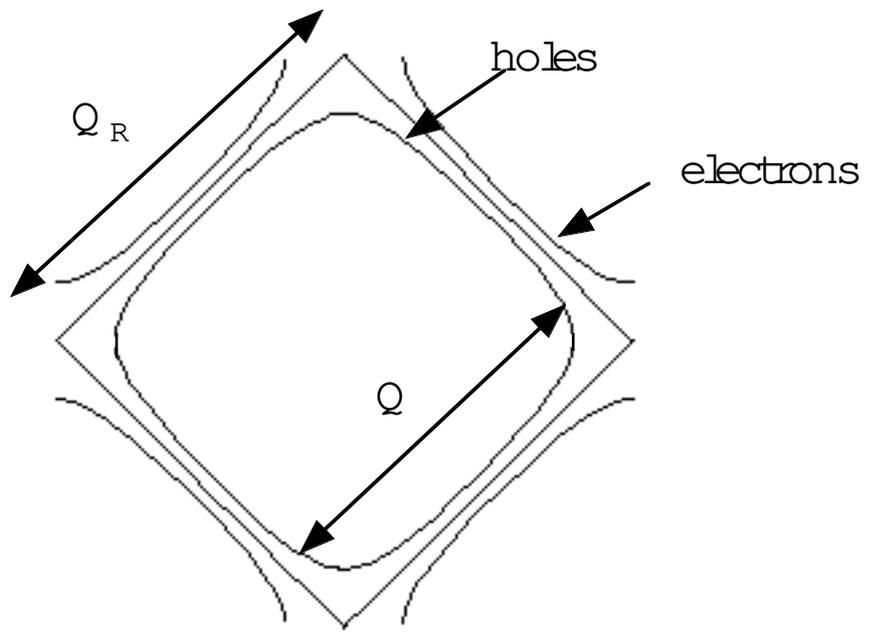
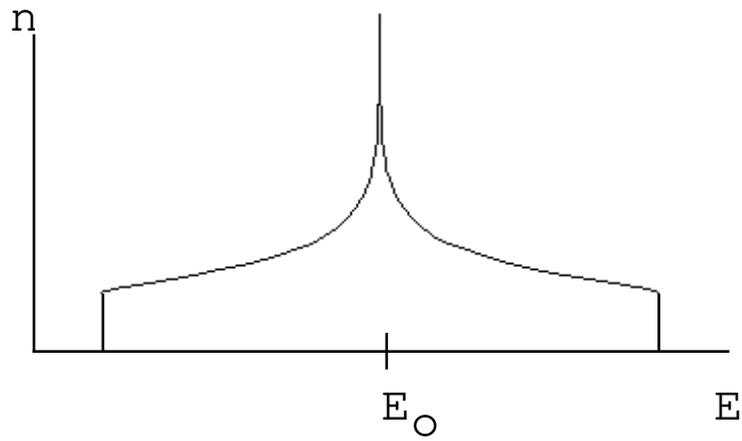
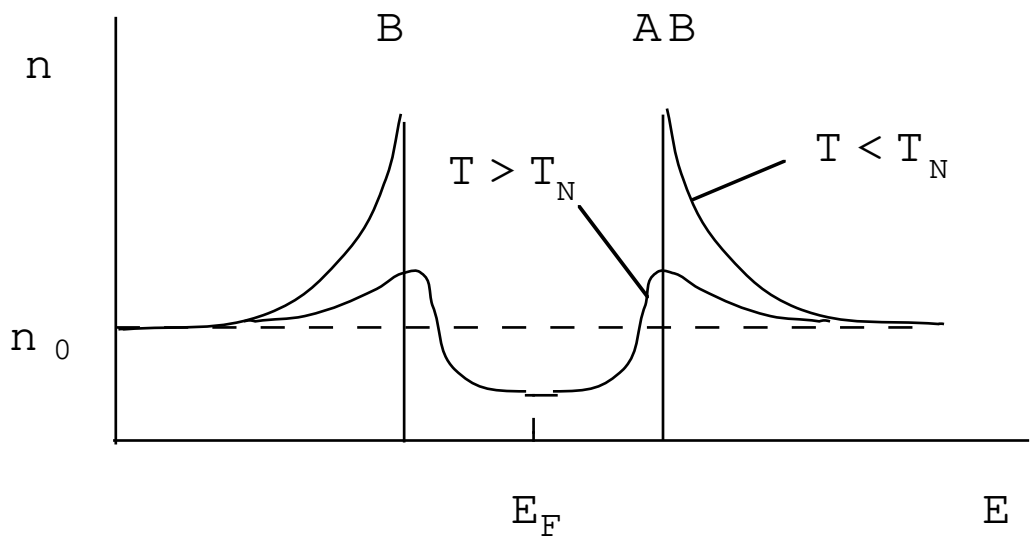
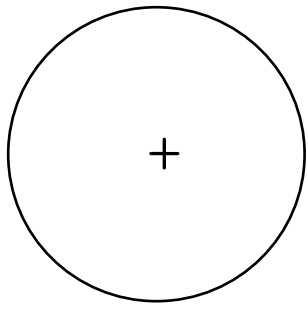
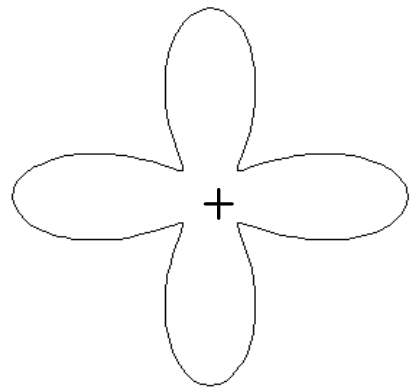
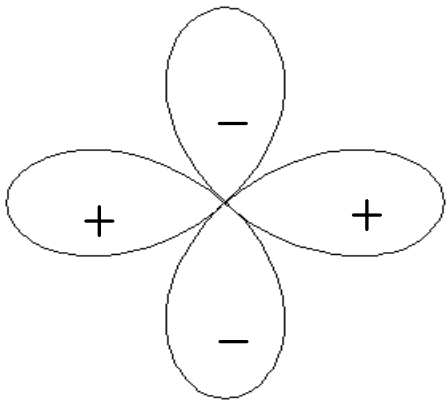
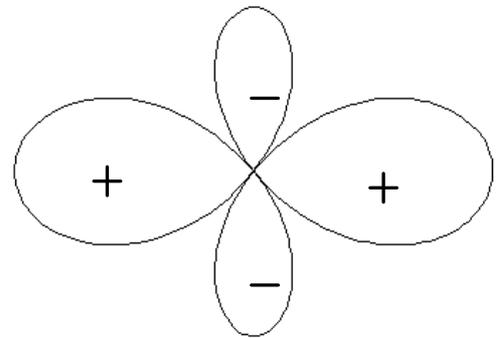
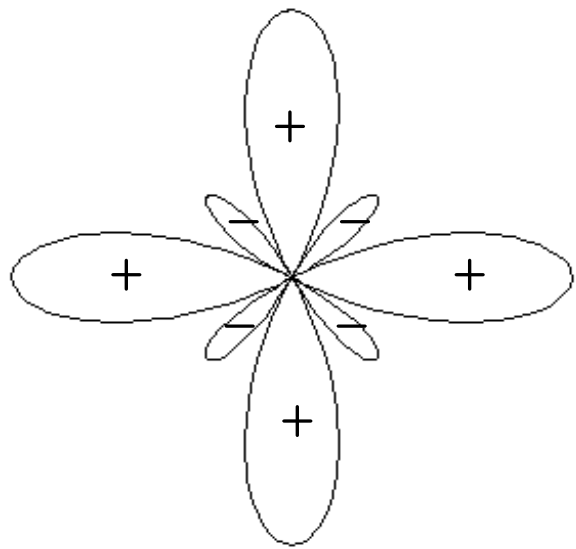
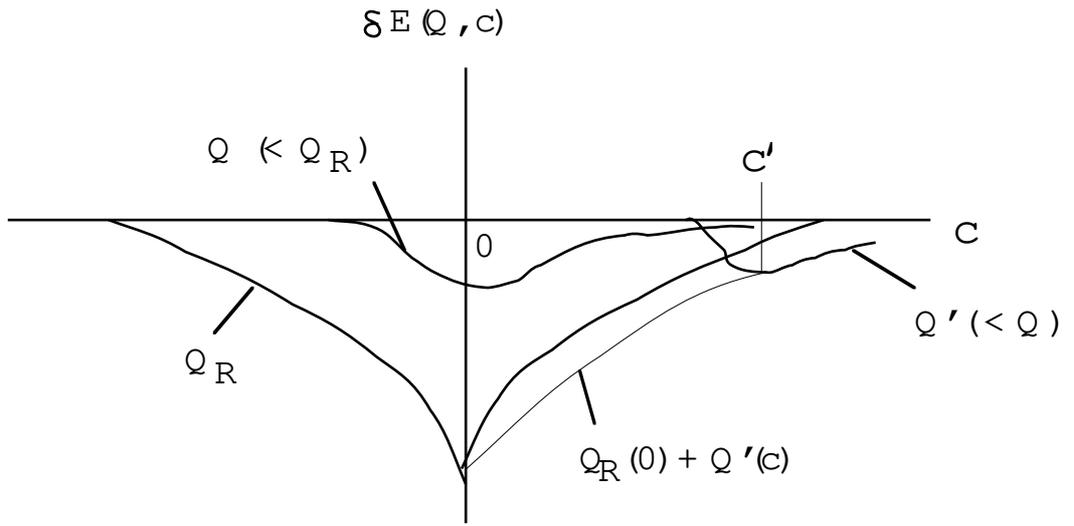
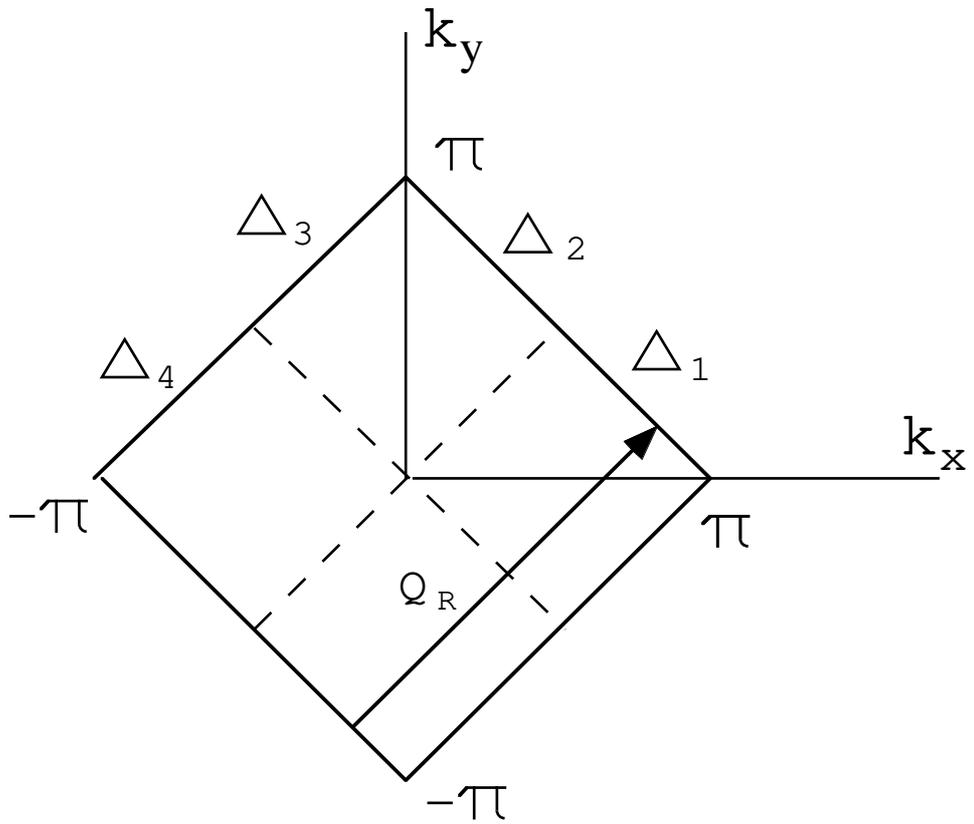
\end{document}